\documentclass[conference]{IEEEtran}
\IEEEoverridecommandlockouts
\usepackage{cite}
\usepackage{amssymb,amsfonts}
\usepackage{amsthm}
\usepackage{textcomp}
\usepackage{xcolor}
\def\BibTeX{{\rm B\kern-.05em{\sc i\kern-.025em b}\kern-.08em
    T\kern-.1667em\lower.7ex\hbox{E}\kern-.125emX}}

\usepackage[]{hyperref}
\usepackage{tabularx}
\usepackage{graphicx}
\usepackage{subcaption}
\usepackage{algorithm}
\usepackage{array}
\usepackage{amsmath}
\usepackage{algpseudocode}
\usepackage{tcolorbox}
\usepackage{lipsum}
\usepackage{multirow}

\usepackage{lineno}

\usepackage{tikz}
\usetikzlibrary{quantikz}
\def\>{\ensuremath{\rangle}}
\def\<{\ensuremath{\langle}}

\newcommand {\Tr} {{\mathrm{Tr}}}
\newcommand {\tr} {{\mathrm{Tr}}}

\newcommand{\nm}[1]{\lVert #1\rVert}

\makeatletter
\renewcommand{\boxed}[2][\fboxsep]{{%
  \setlength{\fboxsep}{#1\fboxsep}\fbox{\m@th$\displaystyle#2$}}}
\makeatother

\newtheorem{theorem}{Theorem}
\newtheorem{lemma}{Lemma}

\usepackage{lineno}

\AtBeginDocument{%
  \providecommand\BibTeX{{%
    Bib\TeX}}}

\begin{document}

\title{Towards Efficient Verification of Computation in Quantum Devices}

\author{
\IEEEauthorblockN{1\textsuperscript{st} Keren Li}
\IEEEauthorblockA{
\textit{College of Physics and Optoelectronic Engineering}\\ 
\textit{Shenzhen University}\\
Shenzhen, China \\
https://orcid.org/0000-0001-5559-0421}
\and
\IEEEauthorblockN{2\textsuperscript{nd} Peng Yan}
\IEEEauthorblockA{\textit{School of Computer Science}\\
\textit{Nanjing University of Posts and Telecommunications}\\
Nanjing, China \\
https://orcid.org/0000-0003-2930-7447}
\and
\IEEEauthorblockN{3\textsuperscript{rd} Hanru Jiang}
\IEEEauthorblockA{\textit{Beijing Institute of Mathematical Sciences and Applications}\\
Beijing, China\\
https://orcid.org/0000-0002-5965-1209}
\and
\IEEEauthorblockN{4\textsuperscript{th} Nengkun Yu}
\IEEEauthorblockA{\textit{Department of Computer Science}\\
\textit{Stony Brook University}\\
Stony Brook, US. \\
https://orcid.org/0000-0003-1188-3032\\
nengkunyu@gmail.com}
}

\maketitle

\begin{abstract}
Designing quantum processors is a complex task that demands advanced verification methods to ensure their correct functionality. However, traditional methods of comprehensively verifying quantum devices, such as quantum process tomography, face significant limitations because of the exponential growth in computational resources. These limitations arise from treating the system as a black box and ignoring its design structure. Consequently, new testing methods must be developed considering the design structure. In this paper, we investigate the structure of computations on the hardware, focusing on the layered interruptible quantum circuit model and designing a scalable algorithm to verify it comprehensively. Specifically, for a given quantum hardware that claims to process an unknown $n$ qubit $d$ layer circuit via a finite set of quantum gates, our method completely reconstructs the circuits within a time complexity of $O(d^2 t \log (n/\delta))$, guaranteeing success with a probability of at least $1-\delta$. Here, $t$ represents the maximum execution time for each circuit layer. Our approach significantly reduces execution time for completely verifying computations in quantum devices, achieving double logarithmic scaling in the problem size. Furthermore, we validate our algorithm through experiments using IBM's quantum cloud service, demonstrating its potential applicability in the noisy intermediate-scale quantum era.
\end{abstract}


\section{Introduction}
Recently, quantum computing has garnered considerable interest from researchers due to its unique quantum properties, such as superposition, entanglement, and noncloning principles. Numerous influential quantum algorithms, including Shor's factoring algorithm \cite{shor}, the hidden subgroup problem \cite{ETTINGER200443}, Grover's search \cite{10.1145/237814.237866}, and HHL algorithm \cite{PRL_HHL}, have been proposed to accelerate computational processes significantly. Advances in quantum hardware~\cite{nature2017, doi:10.1126/science.abe8770, 2019-google-Qsupremacy} indicate that achieving practical quantum supremacy may soon become a reality. With the rapid development of quantum devices, a natural question arises:

\textit{How can engineers ensure the validity of an implementation after Google achieves quantum supremacy within their hardware device \cite{2019-google-Qsupremacy}?}

This question is critical and far from trivial. Given the anticipation that quantum computers will soon become commercially viable and play a crucial role, verifying quantum applications that claim to implement quantum computations will become increasingly important. As these applications enter the market, consumers will need reliable methods to implement these services as claimed. These concerns fall under quantum hardware verification, which has garnered considerable interest, though practical, effective, and reliable methods are still being actively explored.

Classical hardware verification ensures that a hardware design, such as a processor or an integrated circuit, functions correctly and meets specified requirements. This process is critical in hardware development to detect and correct errors, thus validating the design's functionality, performance, and reliability. In contrast, quantum hardware verification differs because of the probabilistic nature of quantum mechanics. In classical hardware, physical tests yield deterministic results with a fixed input, making verification straightforward. However, quantum hardware operates differently: the only way to obtain information about its state is through quantum measurement, which collapses the quantum state into one of many possible outcomes, forming a probabilistic distribution. This inherent probabilistic behavior means that a single measurement provides only partial information about the state. Therefore, specific methods for quantum hardware verification must be designed. 
Simply, one can simulate the calculation on a classical computer and compare the results with the quantum device. It is impractical, as simulations now exceed classical computational capacities~\cite{2019-google-Qsupremacy, kim2023evidence}. Moreover, the current lack of reliable quantum computers complicates the efficient validation of computational outputs, particularly for complex problems such as the sampling problem discussed in~\cite{Hangleiter2019sample,doi:10.1126/science.abe8770}. This challenge becomes even more formidable in a hardware-independent context where details about the hardware are unknown. Consequently, given an unknown quantum hardware, how can we efficiently verify computations performed as expected? Alternatively, can we reveal the architectural information of computation to detect and correct design errors?

\begin{figure}[!ht]
   \centering
   \begin{quantikz}[column sep={0.4cm}, row sep={0.6cm}]
      \lstick[2]{\ket{\text{Input}}} & \lstick[1]{$P$} & \qwbundle \qw & \gate{\text{quantum device}} & \qw & \rstick[2]{\ket{\text{Output}}} \qw \\
      & \lstick[1]{$A$} & \qwbundle \qw & \qw & \qw & \qw
   \end{quantikz}
   \caption{Conventional method for quantum process tomography. Resource consumption, the time cost of generating identical copies, grows exponentially with the number of qubits employed by the principal ($P$) and ancillary ($A$) systems.} 
   \label{fig:pre_method}
\end{figure}
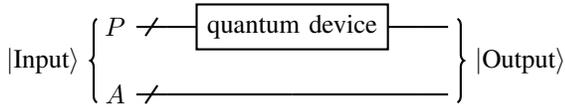

The conventional solution was quantum process tomography (QPT). QPT operates as a black-box model involving quantum measurements on multiple samples to characterize the behavior of a quantum device thoroughly. The standard procedure of QPT is briefly depicted in Fig.~\ref{fig:pre_method}. It involves preparing a maximally entangled input state $\ket{\text{input}}$ between the $n$-qubit principal system $P$ and the $n$-qubit auxiliary system $A$, followed by producing a Choi state $\ket{\mbox{output}}$~\cite{CHOI} for subsequent quantum state tomography.
However, since measurements collapse quantum states, a significant concern arises regarding the sample complexity of quantum state tomography: how many identical copies are required for accurate state reconstruction? The sample complexity is directly linked to the device's execution time to generate identical output copies. Unfortunately, the resource requirements of traditional QPT grow exponentially with the number of qubits, reaching for information-theoretic reasons $O(4^n)$~\cite{10.1145/2897518.2897585}, which hinders its application as a verification tool, especially for many-qubit problems. While shadow tomography~\cite{10.1145/3188745.3188802, kunjummen2021shadow, levy2021classical} represents a recent method aimed at efficiently extracting key information from a quantum state, it does not directly address the comprehensive characterization of quantum device behaviors.

Using the black-box model to test the quantum device significantly limits our operational capabilities.
The exponential cost suggests that a black-box model for quantum device verification is not scalable. In other words, the implementation structure must be investigated to determine if scalable methods are needed to verify quantum hardware.

\textit{In reality, most quantum hardware conducts computations within a layered quantum circuit architecture, meaning the verified target is not an entirely black box. Suppose that we acquire a quantum device from a company employing this architecture. How can we verify the services of the device?}

Moreover, we can stop the device operation at any layer (clarified in Sect.~\ref{sec:circuit_reconstruction}). How can we effectively leverage this control to explore the device? Specifically, the quantum device is structured as a sequence of layers denoted $U_1, U_2, \dots, U_d$, where each layer $U_i$ operates independently and can be treated as a standalone unit. The quantum device can pause at any layer $i$ and output the intermediate state $U_i \cdots U_1 \ket{\text{Input}}$ for further analysis. This capability allows us to decompose the QPT of the entire quantum device into smaller tasks, focusing independently on each layer to improve computational efficiency. This paper introduces an algorithm to perform overlapping state tomography on each layer $U_i$ and efficiently reconstruct the testing unit as $\hat{U}_i$ of a similar function. That is, previously learned layers $\hat{U}_1, \hat{U}_2, \dots, \hat{U}_i$ can be used iteratively to reconstruct the subsequent layer $U_{i+1}$. Given that the quantum device involves a finite universal gate set, we can infer each layer's gate configurations by distinguishing the gate's differences from the Pauli measurements over multiple samples.

In outline, this paper addresses a quantum hardware verification problem by leveraging the layer structure of a quantum circuit, which enables the adaptive recovery of computations executed by these quantum device. Our algorithm can realize a comprehensive verification with a logarithmic time complexity in the number $n$ of qubits and polynomials in the number $d$ of circuit layers.

\subsection*{Key Contribution}
Our study contributes in the following ways:
\begin{itemize}
  \item We formalize the problem associated with verifying computations in quantum devices. We introduce two practical considerations to address this issue as supplementary assumptions for quantum process tomography. Firstly, we assume that the target quantum device exhibits a layered architecture, where the execution time for each layer is upper-bounded by a constant time $t$. Secondly, we assume the ability to interrupt the execution of the quantum device at any layer and capture the corresponding intermediate states as outputs.
  \item We introduce an efficient algorithm for verifying quantum devices. This algorithm facilitates the reconstruction of a target $n$-qubit, $d$-layered quantum circuit with time complexity of $\mathcal{O} (d^2 t \log(n/\delta))$ time, where $1-\delta$ represents the success probability.\footnote{Our focus is solely on the execution time of quantum computations, excluding the post-processing of classical data.} We exploit the layered structure of the target device and employ an adaptive input state preparation procedure. This approach allows us to break down the reconstruction problem of the entire device into quantum process tomography of individual layers. Unlike the standard quantum process tomography, we employ an ``imaginary'' auxiliary system to ``create'' the Choi state for the process tomography of each single layer. Subsequently, the cutting-edge overlapping tomography algorithm is applied to the ``created'' Choi states to improve the efficiency and accuracy of the verification.
  \item We demonstrate our algorithm's capability to eliminate the error accumulation and achieve a high success probability with appropriate sampling. Specifically, the algorithm operates probabilistically, and the measured Choi state exhibits slight fluctuations with varying sample numbers. We demonstrate that this issue can be addressed by minimizing $\epsilon$, a parameter that relates to sample numbers and the minimal distance between gates in a discrete quantum gate set.
  \item We provide experimental demonstrations of our algorithm, showcasing its feasibility through reconstructing a small-scale quantum device programmed with random circuits and quantum Fourier transform (QFT). These experiments were carried out on \textit{ibmq-manila}, a cloud-based noisy quantum processor. As we step into the \emph{noisy intermediate-scale quantum} (NISQ) era~\cite{preskill2018quantum}, characterized by the emergence of devices with a few hundred reliable qubits, the prospect of large-scale quantum computing is becoming increasingly tangible. Although our experiments are constrained by the current scale of available quantum devices, rendering them somewhat preliminary, the results obtained on a real quantum device demonstrate the effectiveness of our approach. This indicates a promising pathway for applying our algorithm in NISQ devices, bolstering their potential in the evolving landscape of quantum computing.
\end{itemize}

\section{Circuit Reconstruction}
\label{sec:circuit_reconstruction}
Consider a quantum device characterized by a quantum circuit $C$.
A critical question naturally arises:

\emph{How can we efficiently learn the entire behavior of the device, i.e., reconstruct the circuit $C$, to verify and validate whether the device behaves correctly? }

We aim to address this inquiry efficiently, reducing the time or operations required for the device to generate the desired output state. We incorporate the practical physical control system into our approach as additional constraints. To simplify the analysis, we introduce and depend on two fundamental assumptions regarding the quantum circuits, culminating in the notion of \emph{layered interruptible circuits}.

\subsection{Layered Interruptible Circuits}
\begin{figure}[!ht]
   \centering
   \begin{quantikz}[column sep={0.3cm}, row sep={0.3cm}]
       \lstick{1} & \gate[4,nwires={3}]{U_1} & \gate[4,nwires={3}]{U_2} & \qw \ \ldots \ & \gate[4,nwires={3}]{U_d} & \qw \\
       \lstick{2} & \qw & \qw & \qw \ \ldots \ \qw & \qw & \qw \\
       \lstick{\vdots} & & & \vdots & & \\
       \lstick{n} & \qw & \qw & \qw \ \ldots \ \qw & \qw & \qw
  \end{quantikz}
  \caption{$d$-layer circuit with $i$-th layer functioning as $U_{i}$.}
  \label{fig:layered_circuit}
\end{figure}
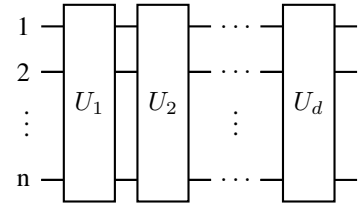

\paragraph{Layered circuits}
First, we assume that the quantum circuit $C$ to be learned is configured with a layered structure. Under this assumption, as shown in Fig.~\ref{fig:layered_circuit}, the circuit $C$ can be described by the equation $C = \prod_{i=d}^1 U_i$, where $U_i$ is the unitary operator describing the behavior of the $i$-th layer, and $d$ denotes the depth of the layers within the circuit $C$. In addition, we also assume that the circuit $C$ is generated from a fixed finite gate set $G = G_1 \cup G_2$, where $G_1$ denotes the set of single-qubit gates and $G_2$ represents the set of two-qubit gates. Each layered circuit can be characterized mathematically as
\begin{align} \label{eq:layered_circuit}
   \textstyle U_i = \bigotimes_{s\in\mathcal{S}} g_{s}
\end{align}
Here, $\mathcal{S}$ is a partition of $[n]$ with each element's size not exceeding $2$, also called the layer structure. The union of all $s\in\mathcal{S}$ is the set $[n]$. $g_{s} \in G$ denotes a gate $g$ applied on the qubits within $s$. $g_{s}$ is either a single-qubit gate ($|s|=1$) or a two-qubit gate ($|s|=2$).

Our assumption finds practical support from the constraints inherent in physical implementation. Quantum computers are commonly configured with a layered circuit architecture that integrates a multitude of single-qubit and two-qubit gates, as outlined in Eq.~\eqref{eq:layered_circuit}. In many physical implementations, realizing two-qubit gates poses greater difficulty and consumes more resources~\cite{kjaergaard2020superconducting,vandersypen2005nmr}. Thus, the two-qubit gates predominantly determine the time required for executing each layer, with $t$ representing the maximum time consumption per layer. Consequently, organizing the operational timeline of a quantum device based on $t$ becomes feasible, facilitating a systematic approach to its utilization and analysis.

\paragraph{Interruptible circuits}
We further assume the capability to pause the quantum device at any layer and measure the intermediate output state. To be specific, one can halt the process at any layer $k \le d$, execute the first $k$ layers, characterized by the unitary  $\prod_{i=k}^1 U_i$, on an input quantum state and subsequently measure the resultant output.

The assumption is essential in the quantum computing field as this capability allows error correction and mitigation strategies to be applied, ensuring the fidelity of computations with external noise and errors. On the other hand, this assumption is firmly rooted in physical reality, where a system can accurately determine processing time. Current quantum computing platforms support sequentially executing multi-layered circuits, one layer at a time, utilizing a programmable pulse generator~\cite{2019-google-Qsupremacy,wu2021strong,alexander2020qiskit}. If we interrupt the device's operation around time $kt$, the resultant output state will likely represent the intermediate state resulting from applying the first $k$ layers.

As a result, the objective is to minimize the time or number of operations the quantum device needs for circuit reconstruction. This involves accessing a quantum device characterized by a layered structure parameterized with $d$ (representing the depth of layers) and $t$ (indicating the maximum time consumption per layer). The task also involves determining the layer structure $\mathcal{S}$ and the corresponding collection of gates $g_{s}$ for all layers.

%The following subsections introduce the key insights that comprise our primary algorithm for solving the \blue{learning} problem.

\subsection{Reconstruct single-layer quantum circuits}
\label{subsec:single_layer}
% First, let us narrow our focus to a single layer $U_i$. The main observation is that reconstructing this layer can significantly increase efficiency, notably reducing the time or operations required by quantum devices. We show that reconstructing a single-layer circuit can be reduced to learning $4$-qubit reduced density matrices of an output state of the circuit. The task is at the cost of sample complexity $\mathcal{O}(\epsilon^{-2} \cdot \log(n /\delta))$, where $\epsilon > 0$ is the allowable deviation and $1-\delta$ the probability. Remarkably, the complexity result is a direct consequence of recent advancements in \emph{quantum overlapping tomography}, a technique used to determine all $m$-qubit reduced density matrices from an $n$-qubit system with a sample complexity $\mathcal{O}(\epsilon^{-2} \cdot 10^m \cdot \log(\binom{n{m}/\delta))$~\cite{2020, yu2020sample}. We refer the readers to \cite{yu2020sample} for details on efficient quantum overlapping tomography due to space limitations. In the following, we describe the procedure for reconstructing a 1-layer circuit.

First, let us narrow our focus to the single-layer quantum circuits.
A basic observation is that we can use \emph{quantum overlapping tomography} to reconstruct the circuit. According to the works~\cite{2020,yu2020sample}, a state tomography of $n$-qubit state can be achieved by overlapping tomography of its $m$-qubit reduced density matrices with the sample complexity 
\begin{equation} \label{eq:complexity}
    \textstyle \mathcal{O}(\epsilon^{-2} \cdot 10^m \cdot \log(\binom{n}{m}/\delta)),
\end{equation}
where $\epsilon$ denotes the trace distance error, and $1-\delta$ represents the confidential level. Readers may refer to these two works for more details about quantum overlapping tomography. The following explains how to use quantum overlapping tomography to reconstruct the single-layer quantum circuit.

By Choi-Jamiolkowski isomorphism, we can fully characterize the properties and behavior of the unitary operation  $U$ as a Choi state. Let $\ket{\Psi}$ be the maximally entangled state between the $n$-qubit principal system $P$ and the $n$-qubit ancillary system $A$,
\begin{equation} \label{eq:n_Bell_states}
   \textstyle \ket{\Psi} = \ket{\Phi}^{\otimes n} = \frac{1}{2^{n/2}}\sum_{i=0}^{2^n-1} \ket{i}_P\ket{i}_A,
\end{equation}
with $\ket{\Phi}$ being the Bell state. Then, the Choi state for the unitary operation is defined as the maximally entangled state between the principal system and the ancillary system,
\begin{equation*}
   \textstyle \ket{\Omega_{U}} = (U \otimes I) \ket{\Psi} = \frac{1}{2^{n/2}}\sum_{i=0}^{2^n-1} (U \ket{i}_P) \otimes \ket{i}_A,
\end{equation*}
where $U$ is applied to the principal system. Notice that the unitary operation $U_i$ is generated from a fixed gate set with only single-qubit and two-qubit gates, as mentioned in Eq.~\eqref{eq:layered_circuit}; thus, the Choi matrix of $U_i$ can be characterized by the tensor product of entangled state involving at most four qubits
\begin{align*}
    \textstyle \ket{\Omega_{U_i}}= \bigotimes_{s\in\mathcal{S}} (g_s\otimes I_s)\ket{\Psi}_s, 
\end{align*}
where $s$ ($|s| \le 2$) denotes the set of qubits that gate $g_s$ is applied to, and $\ket{\Psi}_s = \ket{\Phi}^{\otimes |s|}$ represents the maximally entangled state between $s$ in the principal system and corresponding qubits in the ancillary system.

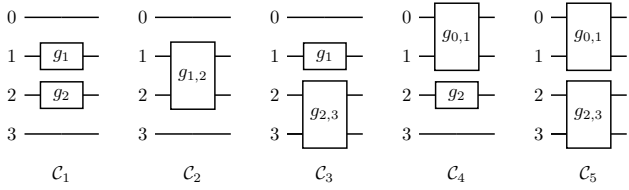
\begin{figure}
   \centering
   \resizebox{0.95\columnwidth}{!}{
   \begin{quantikz}[column sep={0.3cm}, row sep={0.2cm}]
       \lstick{$0$} & \qw & \qw & [0.3cm] \gate[nwires=1,style={draw opacity=0}]{0} & \qw & \qw & [0.3cm] \gate[nwires=1,style={draw opacity=0}]{0} & \qw & \qw & [0.3cm] \gate[nwires=1,style={draw opacity=0}]{0} & \gate[2][0.8cm]{g_{0,1}} & \qw & [0.3cm] \gate[nwires=1,style={draw opacity=0}]{0} & \gate[2][0.8cm]{g_{0,1}} & \qw \\
       \lstick{$1$} & \gate[1][0.8cm]{g_{1}} & \qw & \gate[nwires=1,style={draw opacity=0}]{1} & \gate[2][0.8cm]{g_{1,2}} & \qw & \gate[nwires=1,style={draw opacity=0}]{1} & \gate[1][0.8cm]{g_{1}} & \qw & \gate[nwires=1,style={draw opacity=0}]{1} & & \qw & \gate[nwires=1,style={draw opacity=0}]{1} &  & \qw \\
       \lstick{$2$} & \gate[1][0.8cm]{g_{2}} & \qw & \gate[nwires=1,style={draw opacity=0}]{2} &  & \qw & \gate[nwires=1,style={draw opacity=0}]{2} & \gate[2][0.8cm]{g_{2,3}} & \qw & \gate[nwires=1,style={draw opacity=0}]{2} & \gate[1][0.8cm]{g_{2}} & \qw & \gate[nwires=1,style={draw opacity=0}]{2} & \gate[2][0.8cm]{g_{2,3}} & \qw \\
       \lstick{$3$} &\qw & \qw & \gate[nwires=1,style={draw opacity=0}]{3} & \qw & \qw & \gate[nwires=1,style={draw opacity=0}]{3} & \qw & \qw & \gate[nwires=1,style={draw opacity=0}]{3} & \qw & \qw & \gate[nwires=1,style={draw opacity=0}]{3} & \qw & \qw \\
       & \gate[nwires=1,style={draw opacity=0}]{\mathcal{C}_1} &&& \gate[nwires=1,style={draw opacity=0}]{\mathcal{C}_2} &&& \gate[nwires=1,style={draw opacity=0}]{\mathcal{C}_3} &&& \gate[nwires=1,style={draw opacity=0}]{\mathcal{C}_4} &&&
       \gate[nwires=1,style={draw opacity=0}]{\mathcal{C}_5}
  \end{quantikz}
  }
  \caption{Categories of two-qubit $\{1,2\}$ circuit by the positions of gates. $\mathcal{C}_1$: 1 and 2 are isolated. $\mathcal{C}_2$: 1 and 2 are entangled. $\mathcal{C}_3$ ($\mathcal{C}_4$): 2 (1) is entangled with the remaining qubits. $\mathcal{C}_5$: 1 and 2 are entangled with the remaining qubits.}
  \label{fig:structure_4_qbuits}
\end{figure}

Therefore, all $4$-qubit Choi state $\ket{\Omega_{U_i}}$ can provide adequate information for the reconstruction of this single-layer circuit. Obtaining all $4$-qubit Choi states is accomplished through the overlapping tomography of $4$-qubit $(m=4)$ reduced density matrices.
% In this paper, we employ overlapping tomography of $4$-qubit $(m=4)$ reduced density matrices to learn the Choi state $\ket{\Omega_{U_i}}$. 
We can first figure out the structure $\mathcal{S}$ of layer $\Omega_{U_i}$ from these density matrices. To be specific, Fig.~\ref{fig:structure_4_qbuits} shows the categories of gate positions for $4$-qubit reduced density matrices, where the principal system comprises qubits $\{1, 2\}$, and the corresponding ancillary system is not explicitly shown. Here, qubits $\{0,1,2,3\}$ are introduced for clear illustration, all of which are part of the principal system. Each qubit $i$ is entangled with its corresponding auxiliary qubit $i'$ (not explicitly depicted) in the auxiliary system via the Bell state $\ket{\Phi}_{i, i'}$.

Let $\rho_{1,2}$ be a 4-qubit reduced density matrix corresponding to the register $\{1,2,1',2'\}$, which can also be viewed as a Choi matrix associated with an undetermined 2-qubit quantum process. As shown in Fig.~\ref{fig:structure_4_qbuits}, we can categorize the Choi matrix into one of the five structures.
\begin{equation} \label{eq:Choi_Class}
    \begin{split}
    \mathcal{C}_1 &= \{\ket{\Omega_{g_1}} \otimes \ket{\Omega_{g_{2}}} \mid g_{1}, g_{2} \in G_1 \} \\
    \mathcal{C}_2 &= \{\ket{\Omega_{g_{1,2}}} \mid g_{1,2} \in G_2 \} \\
    \mathcal{C}_3 &= \{\ket{\Omega_{g_1}} \otimes \tr_{3,3'}(\ket{\Omega_{g_{2,3}}}) \mid g_{1} \in G_1, g_{2,3} \in G_2 \} \\
    \mathcal{C}_4 &= \{\tr_{0,0'}(\ket{\Omega_{g_{0,1}}}) \otimes \ket{\Omega_{g_{2}}} \mid g_{0,1} \in G_2, g_{2} \in G_1 \} \\ 
    \mathcal{C}_5 &= \{\tr_{0,0'}(\ket{\Omega_{g_{0,1}}}) \otimes \tr_{3,3'}(\ket{\Omega_{g_{2,3}}}) \mid  g_{0,1}, g_{2,3} \in G_2 \}
    \end{split}
\end{equation}
If the classifications of all 4-qubit reduced density matrices covering the whole system are known, we can infer a single layer's structure $\mathcal{S}$. For example, if a 4-qubit single layer is in the form of $\mathcal{C}_5$ in Fig.~\ref{fig:structure_4_qbuits}, it implies $\mathcal{S} = \{\{0, 1\},\{2, 3\} \}$.

In order to be able to figure out the structure $\mathcal{S}$ and the configuration of gates $\{g_s\}$, it is crucial to ensure that the trace distance error $\epsilon$ is small enough
\begin{align} \label{eq:epsilon}
   \epsilon \le d_{\mathcal{C}} \triangleq \frac{1}{2}\min_{\rho,\rho'\in\mathcal{C}} d(\rho,\rho'), 
\end{align}
such any two different configurations are distinguishable, where $\rho$ and $\rho'$ denote any two distinct 4-qubit reduced density matrices in $\mathcal{C} = \mathcal{C}_1 \cup \mathcal{C}_2 \cup \mathcal{C}_3 \cup \mathcal{C}_4 \cup \mathcal{C}_5$. 
Once the configuration $\mathcal{S}$ and $\{g_s\}$ are established, we can reconstruct the operation $\hat{U}$ for some layer $U$.

For instance, let's consider an ideal Choi matrix $\rho$ associated with a layer, along with its corresponding reconstructed Choi matrix $\widetilde{\rho}$ obtained from state tomography. If the trace distance $d(\Tilde{\rho}, \rho)$ is less than $\epsilon$ defined in Eq.~\eqref{eq:epsilon}, then it can be asserted that $d(\Tilde{\rho}, \rho')> \epsilon$ holds for any $\rho' \in \mathcal{C}$. In such a scenario, the structure $\mathcal{S}$ and the configuration of gates $\{g_s\}$ related to Choi matrix $\rho$ can be accurately inferred. Conversely, if $d(\Tilde{\rho}, \rho)\leq \epsilon$, then there exists the potential for some $\rho' \in \mathcal{C}$ such that $d(\Tilde{\rho}, \rho')< \epsilon$. This situation could lead to inaccurate reconstructions.

%Note that the ancillary qubits in applying Choi-Jamiolkowski isomorphism are introduced to simplify the presentation. Next, we will show that these ancillary qubits can be eliminated without incurring any overhead in Sec.~\ref{subsec:elim-ancilla}.

\subsection{Reconstruct multi-layer quantum circuits}
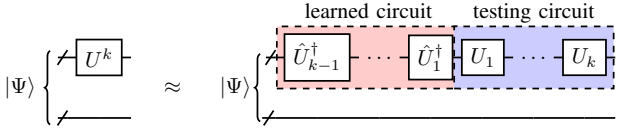
\begin{figure}[!ht]
   \centering
   \resizebox{0.95\columnwidth}{!}{
   \begin{quantikz}[column sep={0.15cm}, row sep={0.6cm}]
      \lstick[3]{\ket{\Psi}} & \qwbundle \qw & \gate{U^k} & \qw & & \midstick[3, brackets=none]{$\approx$} & \midstick[3, brackets=right]{\ket{\Psi}} & \qwbundle \qw & \gate{\hat{U}_{k-1}^\dagger} \gategroup[1,steps=3,style={dashed,fill=red!20, inner xsep=0pt, inner ysep=0pt}, background]{{learned circuit}} & \ \ldots\ \qw & \gate{\hat{U}_{1}^\dagger} & \gate{U_1} \gategroup[1,steps=3,style={dashed,fill=blue!20, inner xsep=0pt, inner ysep=0pt}, background]{{testing circuit}} & \ \ldots\ \qw & \gate{U_k} & \qw \\
      & \qwbundle \qw & \qw & \qw &&&& \qwbundle \qw & \qw & \qw & \qw & \qw & \qw & \qw & \qw
   \end{quantikz}
   }
   \caption{Prepare the Choi state for the $k$-th layer via constructed first $k-1$ layers.}
   \label{fig:adaptive-step}
\end{figure}

As long as we know how to reconstruct a single-layer quantum circuit, we can rebuild a multi-layer quantum circuit by recursive construction. 
The idea comes from the observation that applying the $k$-th layer to the state $\ket{\phi}$ is equivalent to applying the first $k$ layers to the intermediate state $({U}_{k-1}\cdots{U}_{1})^{\dagger}\ket{\phi}$. Consequently, if we have completely learned the first $k-1$ layers, that is, well-estimated $\hat{U}_{1}, \ldots, \hat{U}_{k-1}$, then we can reconstruct the $k$-th layer by the technique described in the preceding subsection. The notation $\hat{U}$ denotes the well-estimated reconstruction of the target circuit $U$. Fig.~\ref{fig:adaptive-step} shows how to prepare the Choi state for reconstructing the $k$-th layer described by $U_k$. The red box denotes the first $k-1$ layers of the reconstructed circuit we learned in previous iterations, and the blue box represents the first $k$ layers of the actual interruptible circuit.

Recall that we assume the execution time of each layer is upper-bounded by $t$. As shown in Fig.~\ref{fig:adaptive-step}, the circuit on the right side for preparing the Choi state has $2k-1$ layers; hence, its execution consumes $(2k-1)t$ time per iteration. Given the need to reconstruct all $d$ layers in Fig.~\ref{fig:layered_circuit}, a single sampling requires a total time of $$\textstyle \mathcal{O}(d^2 t) = \mathcal{O}(\sum_{k=1}^d (2k-1)t).$$ Following  Eq.~\eqref{eq:complexity} of quantum overlapping tomography, 
achieving the tomography of all 4-qubit reduced density matrices with a success probability of $1-\delta$ demands $\mathcal{O}(\epsilon^{-2} \cdot \log(n/\delta))$ samplings,  considering $m=4$ is a constant. Therefore, the total execution time for reconstructing the layered circuit $C$ in Fig.~\ref{fig:layered_circuit} amounts to 
\begin{equation*}
    \mathcal{O}(d^2 \cdot t\cdot\epsilon^{-2} \cdot \log(n/\delta)). 
\end{equation*}

\subsection{Eliminating ancillary qubits using pseudo-measurements}
\label{subsec:elim-ancilla}
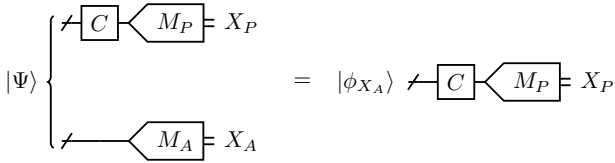
\begin{figure}[!ht]
   \centering
   \resizebox{0.95\columnwidth}{!}{
   \begin{quantikz}[column sep={0.15cm}, row sep={0.3cm}]
      \lstick[3]{\ket{\Psi}} & \qwbundle \qw & \gate{C} & \measuretab{M_P} & \cw & X_P & & \midstick[3, brackets=none]{$=$} & \\
      &&&&&&&& \ket{\phi_{X_A}} \hspace{0.2cm} & \qwbundle \qw & \qw & \gate{C} & \measuretab{M_P} & \cw & X_P \\
      & \qwbundle \qw & \qw & \measuretab{M_A} & \cw & X_A &&&&&&&&&
   \end{quantikz}
   }
   \caption{Basic idea for eliminating ancillary qubits}
   \label{fig:elim-ancilla}
\end{figure}

In the previous subsections, we explained the method for preparing Choi states and detailed the process for reconstructing each layer through the overlapping tomography of all 4-qubit reduced density matrices of the Choi state.
The tomography of 4-qubit reduced density matrices involves performing Pauli measurements randomly on multiple copies. The circuit depicted on the left side of Fig.~\ref{fig:elim-ancilla} illustrates the Pauli measurements $M_P$ and $M_A$ conducted on the Choi state, targeting the principal $(P)$ and ancillary $(A)$ systems. $X_P$ and $X_A$ are the corresponding measurement results.

There is an observation that we do not need the ancillary system to do the tomography as long as an appropriate input state is given to the principal system.
As shown in the circuit on the right side of Fig.~\ref{fig:elim-ancilla}, given the measurement result $X_A$, we can prepare an input state $\ket{\phi_{X_A}}$ to yield the equivalent measurement outcome $X_P$. $\ket{\phi_{X_A}} = \ket{\psi_{X_A}}/\nm{\ket{\psi_{X_A}}}$ is the normalization of $\ket{\psi_{X_A}}$,
\begin{equation*}
    \ket{\psi_{X_A}}\!\bra{\psi_{X_A}} = \tr_A ((I_P\otimes M_{X_A})\ket{\Psi}\!\bra{\Psi}(I_P\otimes M_{X_A}^\dagger)),
\end{equation*}
where $M_{X_A}$ is the corresponding projection for the measurement $M_A$ with measurement outcome $X_A$. Notice that $X_A$ exhibits a uniform distribution over $\{0,1\}^{\otimes n}$ as a result of $\ket{\Psi}$ (Eq.~\eqref{eq:n_Bell_states}) being the tensor product of Bell states, we can generate $M_A$, $X_A$ with a classical random number generator. Subsequently, the state $\ket{\phi_{X_A}}$ can be directly fed into the principal system to replicate the same measurement behavior.
\par Additionally, it's worth noting that $\ket{\phi_{X_A}}$ is a tensor product of the eigenstates of Pauli operators. The preparation of $\ket{\phi_{X_A}}$ requires at most two additional single-qubit gates for each qubit (shown in Algorithm~\ref{alg:prepinit}), introducing merely a constant factor to the algorithm's time complexity. Therefore, we claim these ancillary qubits can be eliminated without incurring overhead.

\section{Verification Algorithm}
\label{sec:main_algorithm}
Our primary findings are summarized in Theorem~\ref{thm:learnmulti} and the main procedure~\textsc{LearnMulti} outlined in Algorithm~\ref{alg:learnmulti}, which incorporates the concepts discussed in Sec.~\ref{sec:circuit_reconstruction}.
Throughout this section, the notation $\boxed[.5]{C}$ denotes an unknown $n$-qubit $d$-layered interruptible testing circuit, and $\boxed[.5]{C(k)}$ represents the testing circuit interrupted at layer $k(k\leq d)$. We enclose circuit $C$ within a box to emphasize that it is unknown.

\begin{theorem}[Efficient Verification Algorithm of Layered Interruptible Quantum Devices by \textsc{LearnMulti}]
Given an unknown $d$-layered interruptible circuit $\boxed[.5]{C}$, with an error bound $\epsilon <\frac{1}{2}\min_{\rho,\rho'\in\mathcal{C}} d(\rho,\rho')$, and $N \ge 2^5\cdot 10^4 \cdot \epsilon^{-2} \cdot \log(2d\binom{n}{2}/\delta)$,
we have $\hat{U}_{C}=\textsc{LearnMulti}$ $(\boxed[.5]{C},N)$, with probability $1-\delta$, requiring $dN$ copy of samples, which completely characterizes the unknown $\boxed[.5]{C}$.
\label{thm:learnmulti} 
\end{theorem}

The proof of Theorem~\ref{thm:learnmulti} follows directly from Lemma~\ref{lm:learnsingle} and~\ref{lm:paulitomo}.
According to Lemma~\ref{lm:paulitomo}, $N \ge 2^5\cdot 10^m \cdot \epsilon^{-2} \cdot \log(2\binom{n}{m}/\delta)$, where $m=4$ and $\delta$ is adjusted to $\delta /d$ in our scenario. 
The term $\binom{n}{2}$ is clarified as follows.
We choose the input state as $\ket{\Phi}_{11'}\otimes \cdots\otimes \ket{\Phi}_{nn'}$, labeling $\{1,\cdots,n\}$ as the principal system and $\{1'\cdots,n'\}$ as the ancillary part, where $\ket{\Phi}$ represents a Bell state.
When considering the principal and ancillary parts within a measured Choi state, they are assessed in pairs.
Thus, it is only necessary to deal with $\binom{n}{2}$ 4-qubit reduced density matrices of qubits $\{i,j,i',j'\}$, instead of all $\binom{2n}{4}$ 4-qubit reduced density matrices.

\begin{algorithm}[ht]
   \begin{algorithmic}[1]
      \Procedure{LearnMulti}{$\boxed[.5]{C}, N$}
         \State $\hat{U}\gets I$
         \For {$k=1$ to $d$} % $K=1$ to $k$
            \State $\hat{U} \gets \textsc{LearnSingle}( \boxed[.5]{C(k)}\cdot\hat{U}^\dagger,N) \cdot \hat{U}$
         \EndFor
         \State\Return $\hat{U}$
      \EndProcedure
   \end{algorithmic}
   \caption{\textsc{LearnMulti}: algorithm for reconstructing layered circuits. Here, $\boxed[.5]{C(k)}\cdot\hat{U}^\dagger$ denotes that $\hat{U}^\dagger$ is applied before $\boxed[.5]{C(k)}$, shown in Fig.~\ref{fig:adaptive-step}.}
   \label{alg:learnmulti}
\end{algorithm}

\begin{algorithm}[ht]
   \begin{algorithmic}[1]
      \Procedure{LearnSingle}{$\boxed[.5]{C'}, N$}
         \For {$i=1$ to $N$}
            \State $(\ket{\phi_i}, M_{A_i}, X_{A_i}) \gets \textsc{PrepInit}(n)$
            \State Uniformly sample $M_{P_i}\in \{\{\frac{1}{2}(\sigma_0 \pm \sigma_j) \}\mid j=1,2,3 \}^{\otimes n}$
            \State Apply $\boxed[.5]{C'}$ to $\ket{\phi_i}$, measure it using $M_{P_i}$, record the measurement result in $X_{P_i}$;
         \EndFor
         \State $\{(s\cup (s+n), \hat{\rho}_{s\cup (s+n)}) \mid s\subseteq[n] \text{ and } |s|=2\} \gets \textsc{PauliTomo}( 4, (M_{P_1} \otimes M_{A_1}, \{X_{P_1}, X_{A_1}\}), \ldots,  (M_{P_N} \otimes M_{A_N}, \{X_{P_N}, X_{A_N}\}))$
         \State $\hat{U}\gets I$; $\mathcal{S} \gets \emptyset$
         \ForAll {$s \subseteq [n]$ and $|s|=2$}
            \State $s'\gets s\cup (s+n)$
            \If {there is $g\in G_2$ s.t. $d\left(\hat{\rho}_{s'}, g\otimes I \ket{\Phi}^{\otimes 2} \right)<\epsilon$}
               \State $\mathcal{S} \gets \mathcal{S} \cup \{ s \}$;
               \State $\hat{U} \gets \hat{U} \cdot g_{s}$
            \EndIf
         \EndFor
         \ForAll {$j\in[n]-\mathcal{S}$}
            \State $s'\gets \{j,j+n\}$
                \State find $\hat{\rho}_{s}$ and $s' \subseteq s$
            \State $\hat{\rho}_{s'} = \tr_{s-s'}(\hat{\rho}_{s})$ 
            \If {there is $g\in G_1$ s.t. $d\left(\hat{\rho}_{s'}, g\otimes I \ket{\Phi} \right)<\epsilon$}
               \State $\hat{U} \gets \hat{U} \cdot g_{\{j\}}$
            \EndIf
         \EndFor
         \State\Return $\hat{U}$
      \EndProcedure
   \end{algorithmic}
   \caption{\textsc{LearnSingle}: algorithm for reconstructing single layer circuit. Here $G = G_1 \cup G_2$ is the gate set, where $G_1$ ($G_2$) is the set of one-qubit (two-qubit) gates and $g_s$ denotes the gate $g$ applied on qubits in $s$. }
   \label{alg:singlelayer}
\end{algorithm}

\begin{lemma}[\textsc{LearnSingle} is sample efficient]
   Given an $n$-qubit single layer circuit $\boxed[.5]{C'}$, and $N\ge 2^5\cdot 10^4 \cdot \epsilon^{-2} \cdot \log(2d\binom{n}{2}/\delta)$.
   If $\epsilon < \frac{1}{2}\min_{\rho,\rho'\in\mathcal{C}} d(\rho,\rho')$, 
   then we have $\hat{U}_{C'}=\textsc{LearnSingle}(\boxed[.5]{C'},N)$ with probability $1-\delta/d$,
   costing $N$ copies of samples, which completely characterizes the unknown single layer $\boxed[.5]{C'}$.
   \label{lm:learnsingle}
\end{lemma}

Within the procedure \textsc{LearnSingle}, we employ the sub-procedure \textsc{PrepInit} detailed in Algorithm~\ref{alg:prepinit} for the pseudo-measurement described in Fig.~\ref{fig:elim-ancilla}.
This step aims to eliminate ancillary qubits initially employed for generating Bell states.
The procedure for preparing the initial state on each qubit $i$ (procedure in the \textbf{for} loop in Algorithm~\ref{alg:prepinit}) is achieved in the classically controlled circuit, as shown below.
\begin{center}
   \begin{quantikz}[column sep={0.3cm}, row sep={0.3cm}]
      & & \\
      \lstick{$\ket{0}$} & \gate{X} & \gate{H} & \gate{S} & \qw & \rstick{$\hspace{-2ex}\ket{\phi_{X_A}}_i$} \\
      & \gate[nwires=1,style={draw opacity=0}]{b_X} \vcw{-1} & \gate[nwires=1,style={draw opacity=0}]{b_H} \vcw{-1} & \gate[nwires=1,style={draw opacity=0}]{b_S} \vcw{-1}
   \end{quantikz}
\end{center}
These gates are controlled by boolean variables $b_X$, $b_H$, and $b_S$, determined by the sampled Pauli measurement $P_i$ and measurement result $X_i$ in the following manner, where notations $P_i$ and $X_i$ are specified in Algorithm~\ref{alg:prepinit}.
\begin{equation}
    \allowdisplaybreaks
    \begin{split}
        b_X =& \ \textstyle (P_i=\frac{1}{2}(\{\sigma_0\pm\sigma_2\})\land X_i=1)\lor (P_i\neq\frac{1}{2} \\
        & \ \textstyle (\{\sigma_0\pm\sigma_2\}) \land X_i=-1) \\
        b_H =& \ \textstyle (P_i\neq\frac{1}{2}(\{\sigma_0\pm\sigma_3\})) \\
        b_S =& \ \textstyle (P_i=\frac{1}{2}(\{\sigma_0\pm\sigma_2\}))
    \end{split}
\end{equation}

\begin{algorithm}[ht]
   \begin{algorithmic}[1]
      \Procedure{PrepInit}{n}
         \State Prepare state $\ket{\phi}=\ket{0}^{\otimes n}$
         \ForAll {$i \in [n]$}
            \State Uniformly sample $P_i\in\{\{\frac{1}{2}(\sigma_0 \pm \sigma_j) \}\mid j=1,2,3 \}$
            \State Uniformly sample $X_i\in\{\pm 1\}$
            \If {$P_i=\frac{1}{2}(\{\sigma_0\pm\sigma_2\})\land X_i=1)\lor 
               (P_i\neq\frac{1}{2}(\{\sigma_0\pm\sigma_2\})\land X_i=-1$}
               \State Apply $X$ gate on the $i$-th qubit
            \EndIf
            \If {$P_i\neq\frac{1}{2}(\{\sigma_0\pm\sigma_3\})$}
               \State Apply $H$ gate on the $i$-th qubit
            \EndIf
            \If {$P_i=\frac{1}{2}(\{\sigma_0\pm\sigma_2\})$}
               \State Apply $S$ gate on the $i$-th qubit
            \EndIf
         \EndFor
         \State \Return $(\ket{\phi}, \otimes_{i=1}^n P_i, \{X_i|1\leq i\leq n\})$
      \EndProcedure 
   \end{algorithmic}
   \caption{\textsc{PrepInit} for pseudo-measurement on ancillary qubits, as described in Sec.~\ref{subsec:elim-ancilla}}
   \label{alg:prepinit}
\end{algorithm}

In addition, we employ an efficient sample algorithm \textsc{PauliTomo} \cite{yu2020sample} for conducting overlapping tomography, as shown in Alorithm~\ref{alg:paulitomo}. This algorithm generates 4-qubit density matrices in Algorithm \ref{alg:singlelayer}. Due to space constraints, readers may refer to ~\cite{yu2020sample} for more details about Algorithm~\ref{alg:paulitomo}. We use The main result presented in Lemma~\ref{lm:paulitomo}.

\begin{lemma}[\textsc{PauliTomo} is efficient~\cite{yu2020sample}]
   If $(P_1,X_1)$, $(P_2,X_2)$, $\ldots$, $(P_N,X_N)$ are random Pauli measurement and corresponding measurement results on an $n$-qubit state $\rho$, 
   and $N \ge 2^5\cdot 10^m \cdot \epsilon^{-2} \cdot \log(2\binom{n}{m}/\delta)$, 
   then with $1-\delta$ probability we will have $d(\rho_s,\hat{\rho}_s)<\epsilon$ for any $(s,\hat{\rho}_s)$ in $\textsc{PauliTomo}(m,$ $(P_1,X_1)$, $(P_2,X_2)$, $\ldots$, $(P_N,X_N))$.
   \label{lm:paulitomo}
\end{lemma}
\begin{algorithm}[ht]
   \begin{algorithmic}[1]
      \Procedure{PauliTomo}{$m,(P_1,X_1), (P_2,X_2), \ldots, (P_N,X_N)$}
         \State \ldots
         \State \Return $\{(s,\hat{\rho}_s)\mid s\subseteq[n] \text{ and } |s|=m\}$
      \EndProcedure
   \end{algorithmic}
   \caption{\textsc{PauliTomo} for overlapping tomography~\cite{yu2020sample}}
   \label{alg:paulitomo}
\end{algorithm}

% \red{This lemma is still valid if we only estimate $r$ different $m$-qubit reduced density matrices. We must only replace the $\binom{n}{m}$ in the $\log$ by $r$.}

\section{Simulations} \label{sec:exp}
\par This section illustrates simulations carried out on a cloud-based quantum computing platform \textit{ibmq-manila}, aimed at showcasing the effectiveness of our algorithm.

\subsection{Testing Circuits}
\par Our testing circuit comprises four qubits $\{q_1, q_2, q'_1, q_2'\}$, which comes from five available superconducting qubits on \textit{ibmq-manila}. Qubits $\{q_1, q_2\}$ function as the principal system, while qubits $\{q_1', q_2'\}$ serve as the corresponding auxiliary system. The structure of the \textit{ibmq-manila}'s quantum chip \cite{ibmq2021} is depicted as follows, where each qubit interact directly with its neighbors.
\tikzstyle{qcir} = [draw,shape=circle,minimum size=0.25cm,inner sep=0.25cm]
\begin{center}
   \resizebox{0.75\columnwidth}{!}{%
    \begin{tikzpicture}[thick]
      \tikzstyle{operator2}=[draw,fill=white, text width=1.1cm, minimum height=4cm] 
      \tikzstyle{operator3}=[draw, dashed, red, text width=7.7cm, minimum height=1.6cm] 
       \node at (2,0.25) (q1) {};
       \node at (3,0.25) {} edge [<-] (q1);
       \node at (2,-0.25) (q1) {};
       \node at (3,-0.25) {} edge [->] (q1);
       \node at (4,0.25) (q1) {};
       \node at (5,0.25) {} edge [<-] (q1);
       \node at (4,-0.25) (q1) {};
       \node at (5,-0.25) {} edge [->] (q1);
       \node at (6,0.25) (q1) {};
       \node at (7,0.25) {} edge [<-] (q1);
       \node at (6,-0.25) (q1) {};
       \node at (7,-0.25) {} edge [->] (q1);
       \node at (8,0.25) (q1) {};
       \node at (9,0.25) {} edge [<-] (q1);
       \node at (8,-0.25) (q1) {};
       \node at (9,-0.25) {} edge [->] (q1);
       \node[qcir] (op221) at (1.5,0) {$q_1'$} ;  
       \node[qcir] (op222) at (3.5,0) {$q_1$} ;
       \node[qcir] (op226) at (5.5,0) {$q_2$} ;
       \node[qcir] (op226) at (7.5,0) {$q_2'$} ;
       \node[qcir] (op226) at (9.5,0) {$q_3$} ;  
       \node[operator3] (op227) at (4.5,0) {} ;  
      \end{tikzpicture}}
\end{center}

\par
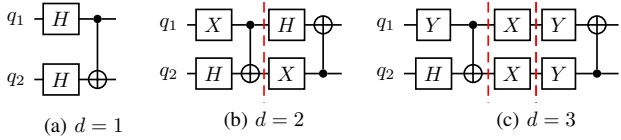
\begin{figure}[!ht]
   \centering
   \resizebox{0.95\columnwidth}{!}{
   \begin{quantikz}[column sep={0.15cm}, row sep={0.5cm}]
      \lstick{$q_1$} & \gate{H} \slice[style ={draw opacity=0}, label style={xshift=0cm, pos=1, anchor=north}]{(a) $d=1$} & \ctrl{1} & \qw \\
      \lstick{$q_2$} & \gate{H} & \targ{} & \qw
   \end{quantikz} \quad
   \begin{quantikz}[column sep={0.15cm}, row sep={0.3cm}]
      \lstick{$q_1$} & \gate{X} & \ctrl{1} \slice[label style={xshift=0cm, pos=1, anchor=north}]{(b) $d=2$} & \gate{H} & \targ{} & \qw \\
      \lstick{$q_2$} & \gate{H} & \targ{} & \gate{X} & \ctrl{-1} & \qw
   \end{quantikz} \quad
   \begin{quantikz}[column sep={0.2cm}, row sep={0.3cm}]
     \lstick{$q_1$} & \gate{Y} & \ctrl{1} \slice{} & \gate{X} \slice{} & \gate{Y} & \targ{} & \qw \\
     \lstick{$q_2$} & \gate{H} & \targ{} & \gate{X} \slice[label style={xshift=0cm, pos=1, anchor=north}]{(c) $d=3$} & \gate{Y} & \ctrl{-1} & \qw
   \end{quantikz}
   }
   \caption{Randomly generated circuits with different $d$.}
   \label{fig:random_circuit}
\end{figure}

 Two tests are conducted to demonstrate the process of reconstructing quantum circuits. The first test is to reconstruct randomly generated circuits from a fixed gate set $\{H, X, Y, Z,$ $CNOT \}$. Fig.~\ref{fig:random_circuit} shows three circuits with pre-set layer depths $d$ and randomly generated quantum gates. Red dashed lines divide the circuits into layers.

\par
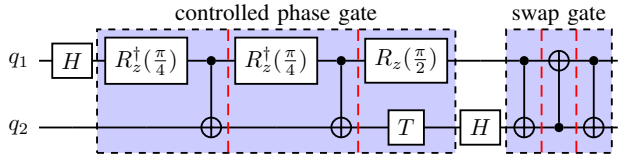
\begin{figure}[!ht]
   \centering
   \resizebox{0.95\columnwidth}{!}{
   \begin{quantikz}[column sep={0.2cm}, row sep={0.4cm}]
      \lstick{$q_1$} & \gate{H} & \gate{R_z^\dagger(\frac{\pi}{4})} \gategroup[2,steps=5,style={dashed,fill=blue!20, inner xsep=0pt, inner ysep=0pt}, background]{{controlled phase gate}} & \ctrl{1} \slice{}& \gate{R_z^\dagger(\frac{\pi}{4})} & \ctrl{1} \slice{} & \gate{R_z(\frac{\pi}{2})} & \qw & \ctrl{1} \slice{}\gategroup[2,steps=3,style={dashed,fill=blue!20, inner xsep=0pt, inner ysep=0pt}, background]{{swap gate}} & \targ{} & \ctrl{1} & \qw \\
      \lstick{$q_2$} & \qw & \qw & \targ{} & \qw & \targ{} & \gate{T} & \gate{H} & \targ{} & \ctrl{-1} \slice{}& \targ{} & \qw
   \end{quantikz}
   }
   \caption{Two-qubit QFT circuit structured with $5$ layers.}
   \label{fig:two_qubit_qft}
\end{figure}

 The second test relates to Quantum Fourier Transformation (QFT) ~\cite{coppersmith}, which is significant to quantum algorithms such as period finding \cite{shor}, HHL algorithm \cite{PRL_HHL} and quantum principal component analysis \cite{Lloyd_2014}. For an $n$-qubit system, QFT is a unitary operation $U$ that maps a computational basis state $\ket{x} = \ket{x_1 x_2 \dots x_{n}}$ $(x_i \in \{0, 1\})$ into a corresponding Fourier basis state $\textstyle U \ket{x} = \ket{\psi_x} = \frac{1}{\sqrt{2^n}} \sum_{y=0}^{N-1} (e^{2\pi i/N})^{x \cdot y} \ket{y}$, where $N=2^n$ and $x \cdot y$ denotes the multiplication between the binary representation of $x$ and $y$. Here, we consider a two-qubit case ($n=2$) mentioned on page 219 in \cite{NI11}, as shown in Fig.~\ref{fig:two_qubit_qft}. This QFT circuit is generated from the gate set $\{H, R_z(\pi/4), R_z(\pi/2), T, CNOT\}$. Notice that the controlled phase gate and the swap gate depicted in \cite{NI11} are decomposed into equivalent sequences of basic gates, as shown in the blue boxes in Fig.~\ref{fig:two_qubit_qft} respectively.

% The two-qubit QFT circuit is generated from the gate set $\{H, R_z(\pi/4), R_z(\pi/2), T, CNOT\}$, where the $z$-rotation gate $R_z(\theta)$ and $T$ gate are
% \begin{eqnarray*}
%    R_z(\theta)=\begin{pmatrix}
%       e^{-i \theta/2}&0\\
%       0&e^{i \theta/2}
%      \end{pmatrix} \qquad
%    T=\begin{pmatrix}
%       1&0\\
%       0&e^{i \pi/8}
%      \end{pmatrix}.
% \end{eqnarray*}

\subsection{Reconstruction}
\par
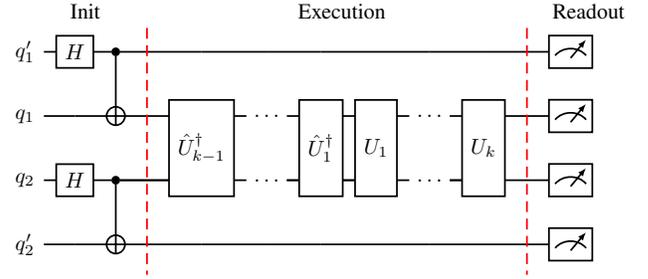
\begin{figure}[!ht]
   \centering
   \resizebox{0.95\columnwidth}{!}{
   \begin{quantikz}[column sep={0.2cm}, row sep={0.4cm}]
      \lstick{$q_1'$} & \gate{H} & \ctrl{1} \slice[label style={xshift=-1cm}]{Init} & [0.5cm] \qw & \qw & \qw & \qw & \qw & \qw \slice[label style={xshift=-3cm}]{Execution} \slice[style ={draw opacity=0}, label style={xshift=1cm}]{Readout} & [0.5cm] \meter{} \\
      \lstick{$q_1$} & \qw & \targ{} & \gate[2, disable auto height]{\hat{U}_{k-1}^\dagger} & \ \ldots \ \qw & \gate[2, disable auto height]{\hat{U}_{1}^\dagger} & \gate[2, disable auto height]{U_1} & \ \ldots \ \qw & \gate[2, disable auto height]{U_k} & \meter{} \\
      \lstick{$q_2$} & \gate{H} & \ctrl{1} & \qw & \ \ldots \ \qw & \qw & \qw & \ \ldots \ \qw & \qw & \meter{} \\
      \lstick{$q_2'$} & \qw & \targ{} & \qw & \qw & \qw & \qw & \qw & \qw & \meter{}
   \end{quantikz}
   }
   \caption{Experimental circuits for reconstructing the $k$-th layer circuit.}
   \label{fig:exp_circuit}
\end{figure}

 According to Algorithm~\ref{alg:learnmulti} and~\ref{alg:singlelayer}, the reconstruction process of the circuit $C$, as shown in Fig.~\ref{fig:layered_circuit}, can be achieved iteratively.
%Specifically, for each iteration, we learn the $U_k$ ($1 \leq k \leq d$), which represents the unitary operation corresponding to the $k$-th layer.
Specifically, learning each layer of the circuit requires the quantum device (\textit{ibmq-manila}) to perform the following three steps, as shown in Fig.~\ref{fig:exp_circuit}.

\begin{enumerate}
   \item \textit{Initialization}. We prepare two pair of two-qubit Bell state $\ket{\Phi} = (\ket{00}+\ket{11})/\sqrt{2}$ over the registers $\{q_1,q_1'\}$ and $\{q_2,q_2'\}$ respectively and ignore the extra unused qubit $q_3$. To be specific, we have the initial state $\ket{\psi_0} = \ket{\Phi}_{1,1'} \otimes \ket{\Phi}_{2,2'}$, where the subscript $i \in \{1, 1', 2, 2'\}$ denotes that a state is over the corresponding qubit $q_i$. %The initialization is realized by applying Hadamard gates and CNOT gates on $\ket{0}^{\otimes 4}$, as shown in Fig.~\ref{fig:exp_circuit}.
   \item \textit{Testing}. Given a testing circuit of $d$ layers,  for the $k$-th ($1\leq k\leq d$) iteration of learning $k$-th layer relating to $\hat{U}_{k}$, a modified circuit, mathematically represented by $\textstyle (\prod_{i=k}^{1} U_i) \cdot (\prod_{i=1}^{k-1} \hat{U}_{i}^{\dagger})$, is applied on qubits $\{q_1, q_2\}$, as shown in Fig.~\ref{fig:exp_circuit}. The operation $\prod_{i=1}^{k-1}\hat{U}_{i}^{\dagger}$ denotes the reconstructed circuit that have been learned from previous $k-1$ iterations, while the operation $\prod_{i=k}^{1} U_i$ represents the first $k$ layers of the actual interruptible testing circuit.
   \item \textit{Readout}. Tomography is employed to registers $\{q_1,q_1'\}$ and $\{q_2,q_2'\}$ with each quantity measured using 8192 shots. A  ``shot'' refers to executing the quantum device once. Due to stochastic noise, multiple shots are necessary to get the expected values for measurement outcomes. The measurements cover the complete Pauli set $\{\sigma_0, \sigma_1, \sigma_2, \sigma_3\}^{\otimes 2}$, which capture the full information of two-qubit density matrices.
\end{enumerate}

Remarkably, we do not directly perform the overlapping tomography for the 4-qubit density matrices as discussed in Sec.~\ref{sec:circuit_reconstruction}. 
Since we have prior knowledge that the available two-qubit gates are restricted to the CNOT gates, only $\mathcal{C}_1$ and $\mathcal{C}_2$ from $\mathcal{C}$ in Eq.~\eqref{eq:Choi_Class} are present in our testing. 
Furthermore, due to limitations in the total number of shots and experimental time on \textit{ibmq-manila}, we instead choose to capture the 2-qubit reduced density matrices, which will not compromise the reconstruction process. 
Consequently, we use $\rho^1_{1, 1'}$ and $\rho^1_{2, 2'}$ to denote the reconstructed 2-qubit reduced density matrices from Pauli measurements. 
Once entanglement is identified in $\rho^1_{1, 1'}$ ($\rho^1_{2, 2'}$), we can disentangle them by applying the CNOT gate, which results in $\rho^2_{1, 1'}$ ($\rho^2_{2, 2'}$). Following this, we can deduce the configuration of local single-qubit gates from the processed states $\rho^2_{1, 1'}$ and $\rho^2_{2, 2'}$.

The deduction of the configuration for single-qubit gates involves classical computations.
Specifically, we aim to identify a single qubit $U_l$ from the predefined fixed gate set to minimize the residual error given by $R= 1- \bra{\Phi} U_{l}^{\dagger} \rho U_{l} \ket{\Phi}$, where $\rho \in \{\rho^2_{1, 1'}, \rho^2_{2, 2'}\}$, $U_{l}$ is applied to either $q_1$ or $q_2$, $\ket{\Phi}$ is the Bell state. Once $U_l$ is determined, the desired reconstructed gate $\hat{U}$ is expressed as $U_c \cdot U_l$. The gate $U_c$ is the identity matrix if the purify of states $\rho^1_{1, 1'}$ and $\rho^1_{2, 2'}$ is close to 1; otherwise, $U_c$ is the CNOT gate.
%Repeating the three steps for each layer can reconstruct the testing circuits depicted in Fig.~\ref{fig:random_circuit} and \ref{fig:two_qubit_qft}.

\subsection{Results}
% Fig.~\ref{R1_exp} and Fig.~\ref{QFT1_exp} exhibit the visualization of the Choi matrices $\rho^1_{1, 1'}$ and $\rho^1_{2, 2'}$ for each layer's reconstruction of the circuits in Fig.~\ref{fig:random_circuit} and Fig.~\ref{fig:two_qubit_qft}. The real and imaginary parts of the $4\!\times\!4$ complex matrices $\rho^1_{1, 1'}$ and $\rho^1_{2, 2'}$ are plotted. As mentioned before, the presence of a CNOT gate can be determined via purity. Specifically, if the purity $\Tr((\rho^1_{1,1})^2)$ and $\Tr((\rho^1_{2,2})^2)$ is closed to $1$, it indicates no entanglement between registers $\{1,1'\}$ and $\{2,2'\}$, suggesting the absence of a CNOT gate between qubits $1$ and $2$, and vice versa.

\begin{figure}[!ht]
  \begin{center}
    \includegraphics[width=0.75\columnwidth]{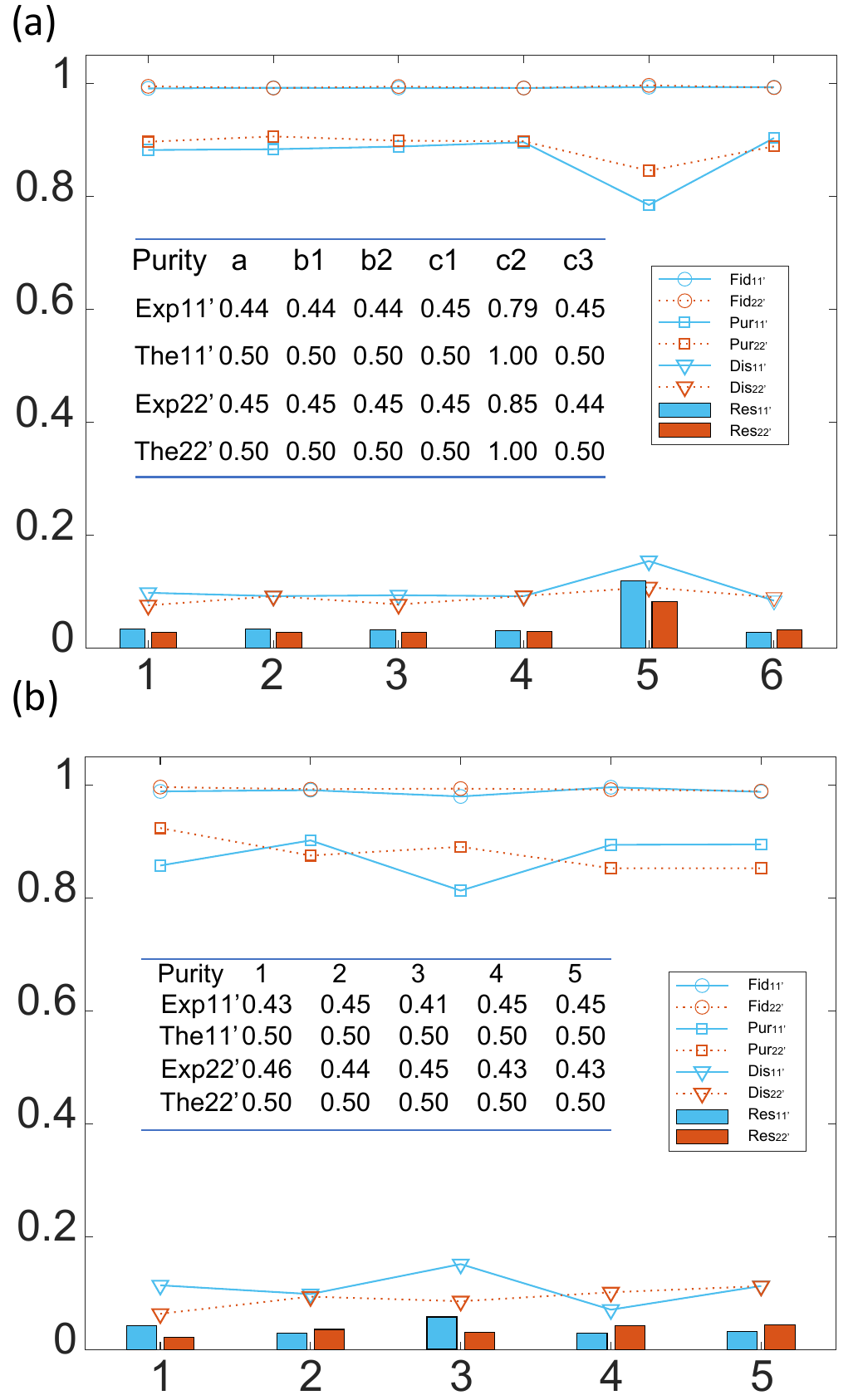}
     \caption{Statistics of $\rho^1_{1,1'}$ and $\rho^1_{2,2'}$ for every layer. (a) is for the random circuit, and (b) for the QFT circuit. Pur${_{ij}}$ denotes the purity of state over qubits $\{q_i,q_j\}$, with both experimental and theoretical values listed in the tables. Fid${_{ij}}$ and Dis${_{ij}}$ are fidelity and distance between experimental and theoretical states over qubits $\{q_i,q_j\}$. Res${_{ij}}$ represents the residual error for the search of local gates.}
     \label{res_exp}
     \end{center}
\end{figure}

First, we obtain all information of the Choi matrices $\rho^1_{1, 1'}$ and $\rho^1_{2, 2'}$ for each layer's reconstruction. We specify them in the appendix. As mentioned before, the presence of a CNOT gate can be determined via purity. Specifically, if the purity $\Tr((\rho^1_{1,1})^2)$ and $\Tr((\rho^1_{2,2})^2)$ is closed to $1$, it indicates no entanglement between registers $\{1,1'\}$ and $\{2,2'\}$, suggesting the absence of a CNOT gate between qubits $1$ and $2$, and vice versa.
The purity for each layer in both tests is presented in the tables of Fig.~\ref{res_exp}. Consequently, it is observed that the CNOT gate is present in every layer except for the second layer of circuit $(c)$. This is consistent with the testing circuit illustrated in Fig.~\ref{fig:random_circuit}. The line chart in Fig.~\ref{res_exp} displays the relative values $\text{Pur}_{11}$ and $\text{Pur}_{22'}$, which is the experimental purity divided by the theoretical value. Parameters $\text{Pur}_{11}$ and $\text{Pur}_{22'}$ approaching $1$ indicate higher accuracy and fidelity in physical implementation.

After determining the placement of CNOT gates, which the structures $\mathcal{S}$ of the layer, we also need to deduce the single-qubit gates from $\rho^2_{1, 1'}$ and $\rho^2_{2, 2'}$, which we get and plot in the appendix.
By conducting a direct search within a predefined gate set, we identify the local quantum gates that minimize the residual error $\text{Res}_{ii'}$, as shown in the histogram in Fig.~\ref{res_exp}.

\par At last, to validate our testing experiments, we compare reconstructed Choi matrices $\rho^1_{1,1'}$ and $\rho^1_{2,2'}$ with their theoretical counterparts, focusing on key metrics such as \emph{relative fidelity} and \emph{distance}.
Notice that the formula employed for the relative fidelity of mixed states in this context is $\tr ({\rho \rho' }) /\sqrt{\tr ({\rho \rho })\tr ({\rho' \rho' })}$. From this standpoint, it can be deduced whether the imperfection is caused by decoherence or not~\cite{vandersypen2005nmr,fortunato2002design}.
These detailed data for all layers are presented in Fig.~\ref{res_exp}. The relative fidelity (distance) values presented in Fig.~\ref{res_exp} are close to 1 (0), indicating that the reconstructed states closely match their theoretical counterparts.

\par In summary, we have demonstrated our algorithm over two kinds of testing circuits on \textit{ibmq-manila}. 
The hardware is designed to deploy quantum gates layer by layer. Notably, the duration of two-qubit quantum gates (180 ns) is much longer than that of single-qubit gates (20ns)~\cite{kandala2021demonstration}, which supports our assumption of an interruptible layered circuit. Given that we do not have full access to every function of the hardware, we cannot control the device by time. Instead, we manage it by layers.
Even though the configurations of our testing circuits are simple because of the device's limitations, this endeavor still stands as a significant proof of concept.
% It lays the groundwork for further exploration and application on larger-scale quantum devices. 
Besides, we want to highlight that Eq.~\eqref{eq:epsilon} assures accuracy, paving the way for future investigations into scenarios with less accuracy.
We also provide a numerical simulation result in the appendix to demonstrate our algorithm's effectiveness.
% In a word, our results of experimental simulations offer preliminary validation for the feasibility of our algorithm, particularly in the context of quantum noise.

\section{Discussions \& Related Work}

For an $n$-qubit quantum device with $d$ layers, our algorithm can reconstruct the circuit with the time complexity $\mathcal{O}(d^2 \cdot t\cdot\epsilon^{-2} \cdot \log(n/\delta))$, achieving a success probability of $1-\delta$, which is guaranteed by method of quantum overlapping tomography in \cite{2020,yu2020sample}. Its feasibility is demonstrated by some experimental simulations over multi-layer circuits supported by a cloud-based quantum platform.

Our method currently relies on two key idealization. First, we require a layered circuit in which each qubit participates in at most one two-qubit gate per layer drawn from a fixed, discrete gate set.
Second, we do not yet model noise: the analysis assumes that the classically learned inverse gates can be re-applied with negligible error, so cumulative error build-up is ignored.
These simplifications leave three open directions. 
The potential for generalizing our approach to accommodate a continuous gate set is exciting. In our current study, the testing circuits are generated from a fixed gate set, wherein the distances between any two distinct quantum gates are finitely enumerable. Thus, there is an expectation for extending this approach to encompass a continuous gate set. Another important goal is establishing theoretical guarantees for our verification method when applied to noisy intermediate-scale quantum devices. Such theoretical foundations would bolster our experimental findings and guide further exploration of noisy quantum devices. Furthermore, experimental demonstrations are still lack as our limited experimental capacity. A compelling challenge lies in convincingly demonstrating the practicability of our method on a large-scale quantum device which requires our no ancilla strategy. This endeavor necessitates collaboration with experimental scientists to validate the method's effectiveness in a broader and more complex context.

Compared to other existing methods, 
such as compressed sensing quantum process tomography~\cite{gross2010quantum}, direct fidelity estimation~\cite{flammia2011direct}, randomized benchmarking~\cite{knill2008randomized} and cross-entropy benchmarking~\cite{boixo2018characterizing}, they have been proposed to enhance efficiency. Specifically, compressed sensing quantum process tomography simplifies the characterization of quantum processes by exploiting sparsity, significantly reducing the required measurements. While it proves advantageous for larger quantum systems, its efficacy relies on the assumption of sparsity. Direct fidelity estimation evaluates a quantum process's fidelity using fewer resources than full QPT. Although convenient for benchmarking and error mitigation, it provides a narrower perspective on the quantum process, yielding only a limited set of metrics. Randomized benchmarking assesses quantum gate performance by measuring how fidelity decays when applying random gate sequences. It offers average error rates rather than detailed process descriptions. Cross entropy benchmarking uses random quantum programs to determine the fidelity of a wide variety of circuits. 
Beyond these four baselines, several newer protocols deserve mention.  
Gate-set tomography~\cite{Nielsen2021gatesettomography} yields a  self-consistent, high-precision model of an entire one- or two-qubit gate set.  Its log-likelihood maximization, however, scales exponentially with the size of the Hilbert space, so present implementations are confined to a few qubits.  
Cycle benchmarking~\cite{Erhard:2019cxk} generalizes randomized benchmarking to multi-qubit ``cycles,” returning SPAM-independent per-cycle process infidelities.  Each cycle is nevertheless reduced to a single scalar and the protocol assumes that the inverse of every cycle (or an efficiently implementable approximation) is available.  
Classical-shadow tomography for quantum channels~\cite{Levy2024Classical} can estimate a wide family of linear functionals with only polylogarithmic sample overhead.  Yet current realizations rely on entangled inputs and the channel–state isomorphism, and the classical post-processing cost still grows rapidly with both circuit depth and the number of measured observables.  
Therefore, these methods are either not generally applicable to practical scenarios or can extract only limited information from the system. In contrast, our approach, based on reasonable assumptions, presents an efficient means of comprehensively understanding the behavior of unknown computations in quantum devices.

In contrast, our overlap-tomography-based verifier strikes a middle ground: it returns a \emph{complete, human-readable reconstruction of every layer} (hence pin-pointing faulty gates), while requiring only $\mathcal O(d^{2}t\log(n/\delta))$ samples—poly-logarithmic in qubit count and polynomial in depth—under the modest assumptions spelled out in Section 2.  Table 2 summarises these trade-offs quantitatively.

\section{Appendix}
\subsection{Quantum Simulation}
\begin{figure}[!ht]
   \centering
   \includegraphics[width=0.95\columnwidth]{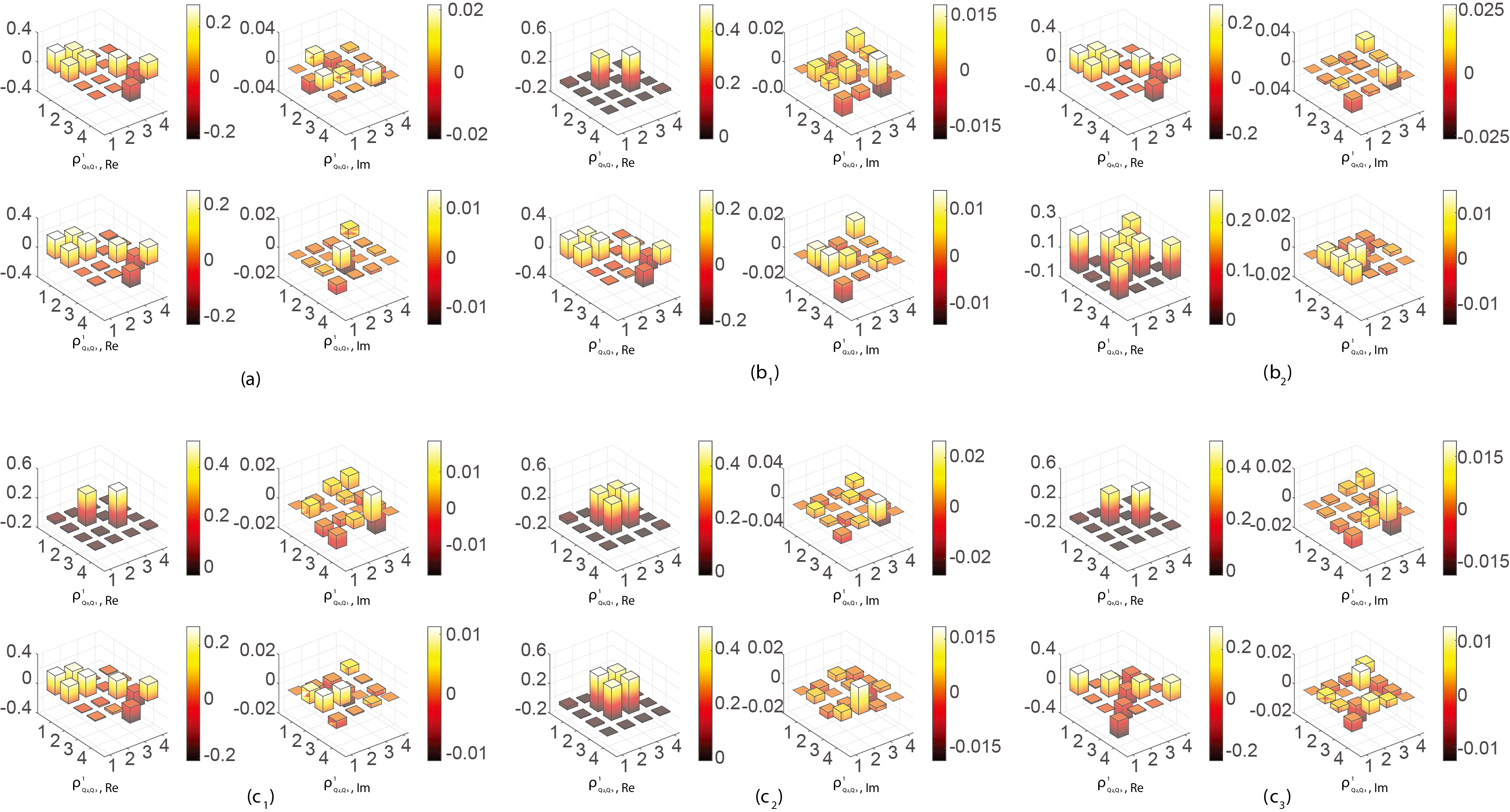}
   \caption{Real and imaginary parts of $\rho^1_{1,1'}$ and $\rho^1_{2,2'}$ for the reconstruction of randomly generated circuits at every layer in Fig.~\ref{fig:random_circuit}. Panels $(a)$, $(b_k)$, and $(c_k)$ refer to circuits illustrated in Fig.~\ref{fig:random_circuit}. The subscript $k \in \{1,2,3\}$ denotes the result from the $k$-th layer.}
   \label{R1_exp}
\end{figure}

\begin{figure}[!ht]
   \centering
   \includegraphics[width=0.95\columnwidth]{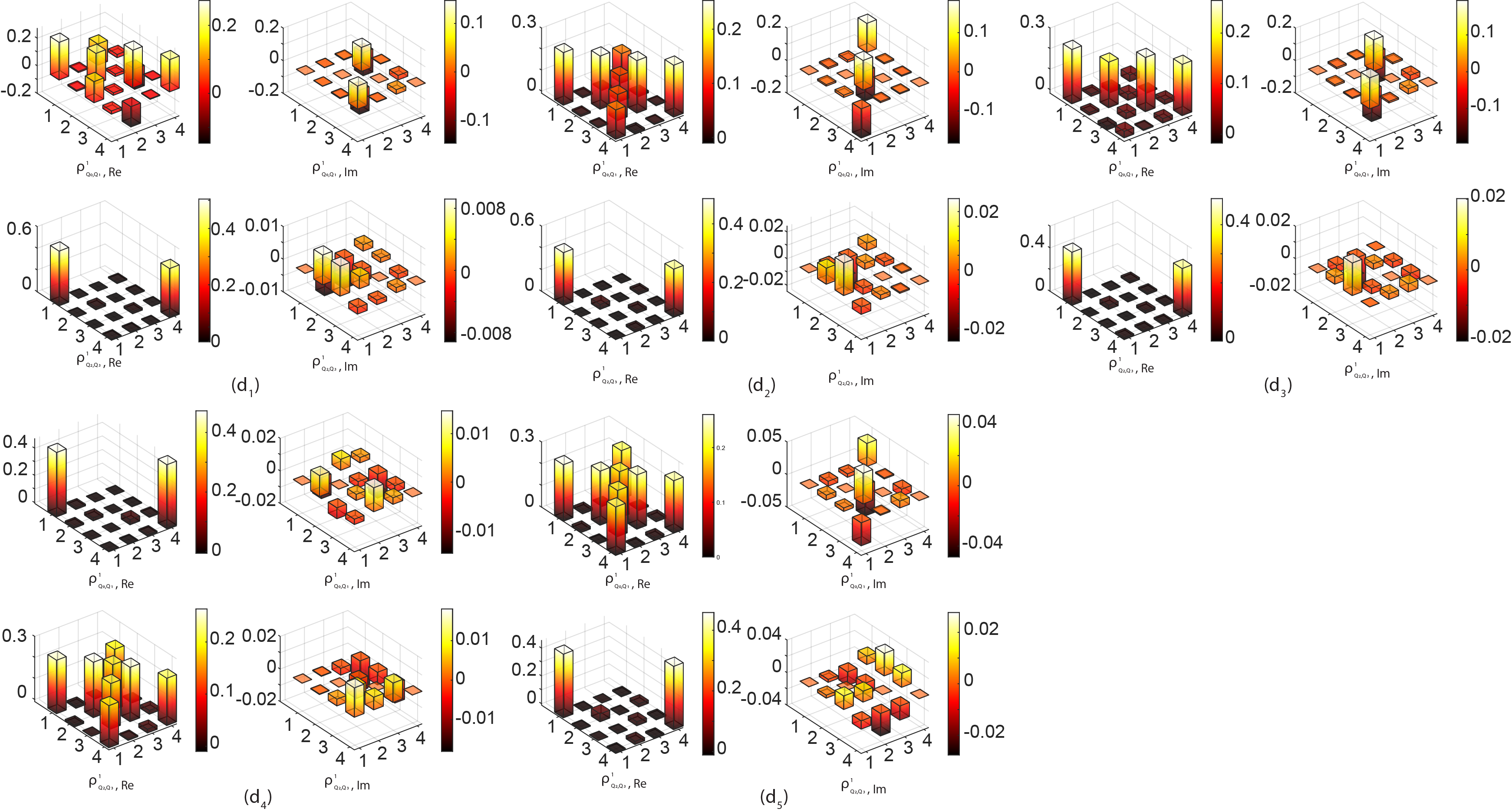}
   \caption{Real and imaginary parts of $\rho^1_{1,1'}$ and $\rho^1_{2,2'}$ for the reconstruction of the two-qubit QFT circuits at every layer in Fig.~\ref{fig:two_qubit_qft}. The subscript $k \in \{1,2,3,4,5\}$ in $(d_k)$ denotes the result from the $k$-th layer.}
   \label{QFT1_exp}
\end{figure}

Fig.~\ref{R1_exp} and Fig.~\ref{QFT1_exp} exhibit the visualization of the Choi matrices $\rho^1_{1, 1'}$ and $\rho^1_{2, 2'}$ for each layer's reconstruction of the circuits in Fig.~\ref{fig:random_circuit} and Fig.~\ref{fig:two_qubit_qft}. The real and imaginary parts of the $4\!\times\!4$ complex matrices $\rho^1_{1, 1'}$ and $\rho^1_{2, 2'}$ are plotted. 

\begin{figure}[!ht]
   \centering
   \includegraphics[width=0.95\columnwidth]{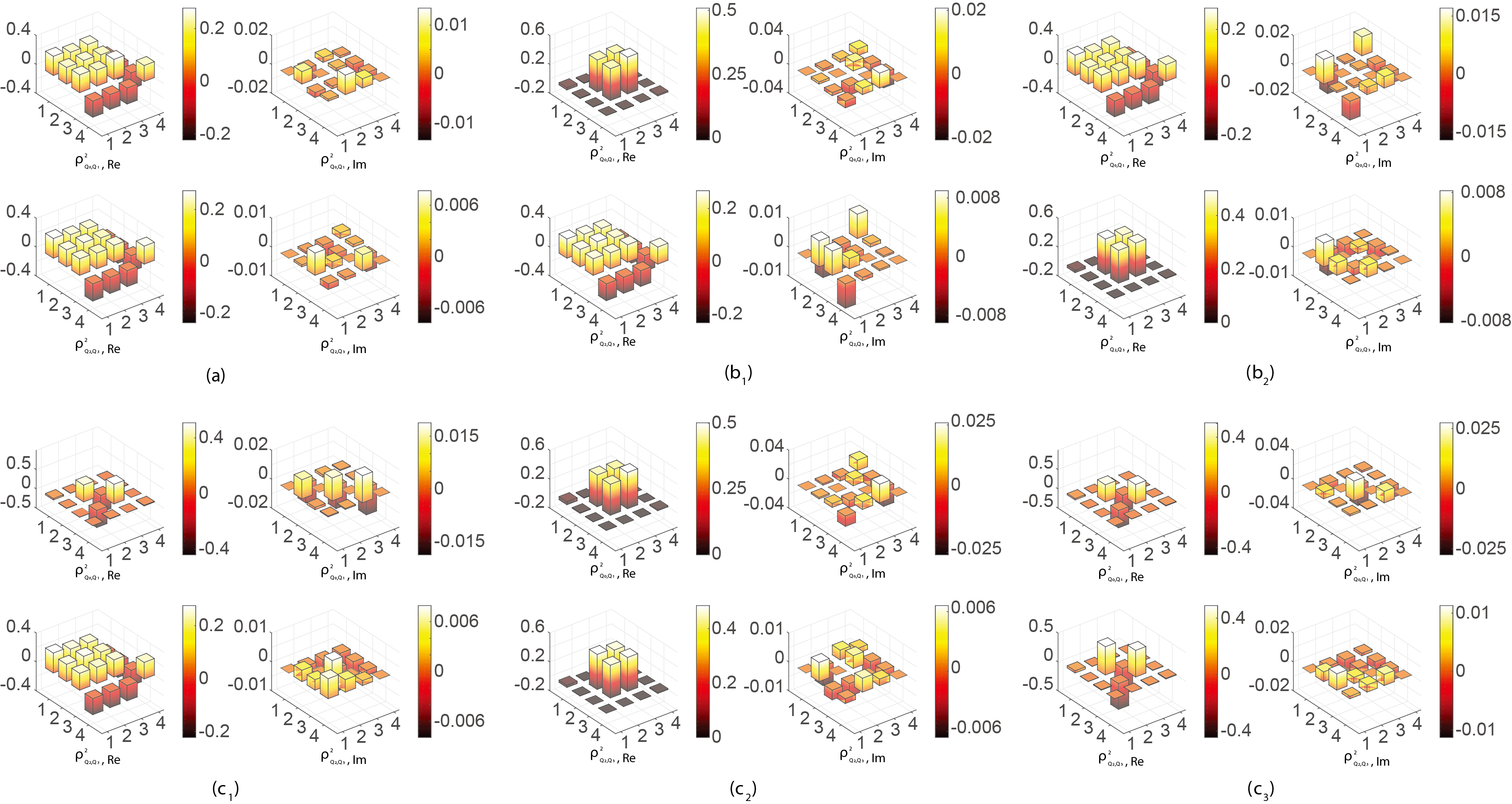}
   \caption{ Real and imaginary parts of $\rho^2_{1,1'}$ and $\rho^2_{2,2'}$ for the reconstruction of randomly generated circuits at every layer in Fig.~\ref{fig:random_circuit}. Panels $(a)$, $(b_k)$, and $(c_k)$ refer to circuits illustrated in Fig.~\ref{fig:random_circuit}. The subscript $k \in \{1,2,3\}$ denotes the result from the $k$-th layer.}
   \label{R2_exp}
\end{figure}

\begin{figure}[!ht]
   \centering
   \includegraphics[width=0.95\columnwidth]{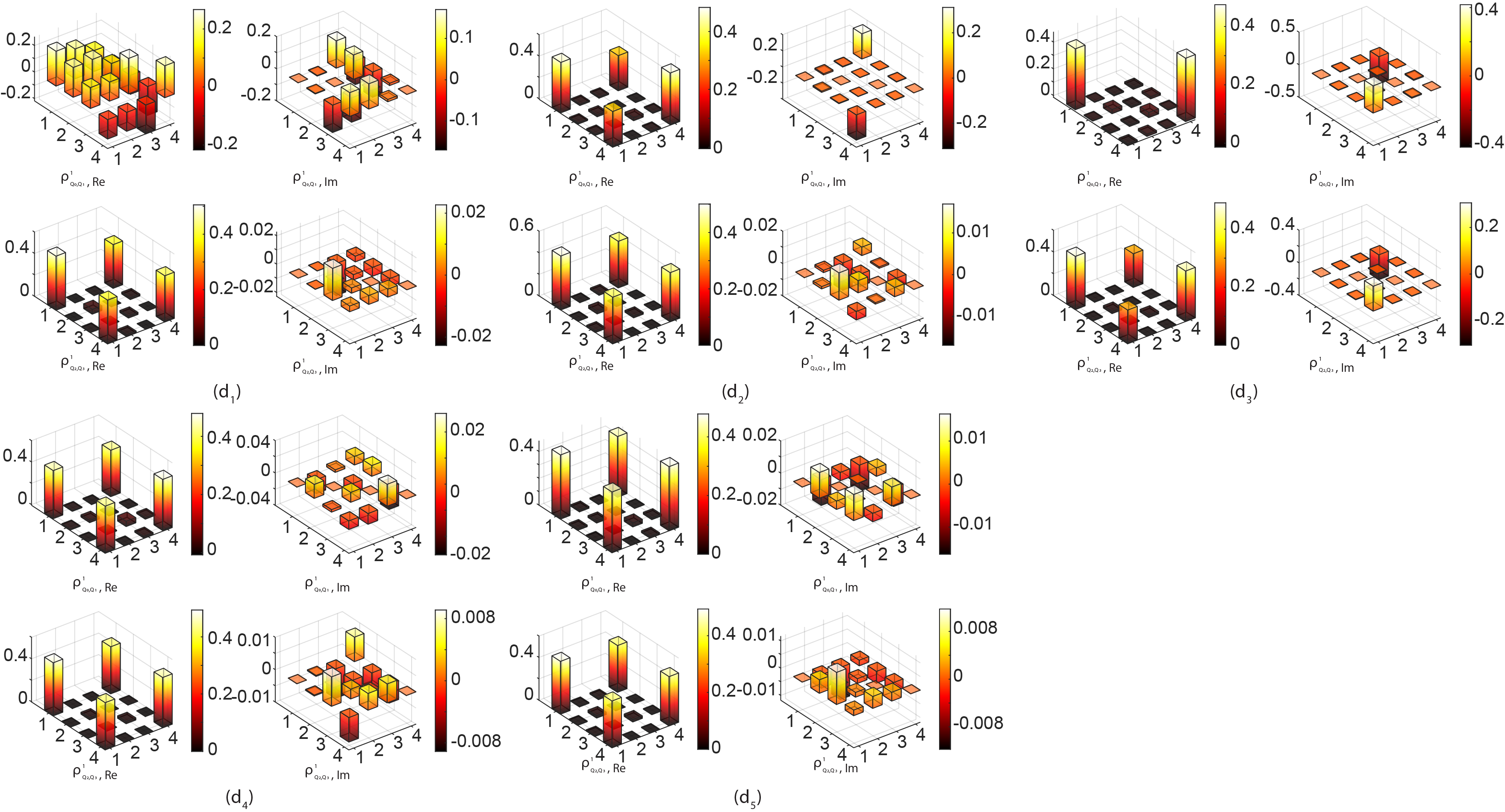}
   \caption{Real and imaginary parts of $\rho^2_{1,1'}$ and $\rho^2_{2,2'}$ for the reconstruction of the two-qubit QFT circuit at every layer in Fig.~\ref{fig:two_qubit_qft}. The subscript $k \in \{1,2,3,4,5\}$ in $(d_k)$ denotes the result from the $k$-th layer.}
   \label{QFT2_exp}
\end{figure}

$\rho^2_{1, 1'}$ and $\rho^2_{2, 2'}$ are illustrated in Fig.~\ref{R2_exp} and Fig.~\ref{QFT2_exp}. As mentioned before, $\rho^2_{1, 1'}$ and $\rho^2_{2, 2'}$ are generated by applying either the CNOT gate or the identity gate to state $\rho^1_{1, 1'}$ and $\rho^1_{2, 2'}$. For instance, $\rho^2_{i, i'}$ and $\rho^1_{i, i'}$ ($i=1,2$) exhibit high similarity in scenarios where the CNOT gate is absent in the second layer of the circuit $(c)$ in Fig.~\ref{fig:random_circuit}.

\subsection{Classical Simulations}
Due to the resource constraints imposed by the platform, we chose tomography of a 2-qubit rather than a 4-qubit reduced density matrix in our experiments. The fundamental justification behind this choice is that the CNOT gate acts as the exclusive source of entanglement within the principal system, and the system's size is confined to merely four qubits. In this subsection, we perform some classical simulations to demonstrate the effectiveness of our algorithms discussed in Sec.~\ref{sec:circuit_reconstruction} and Sec.~\ref{sec:main_algorithm}.

\begin{figure*}[!ht]
  \centering
  \includegraphics[width=2\columnwidth]{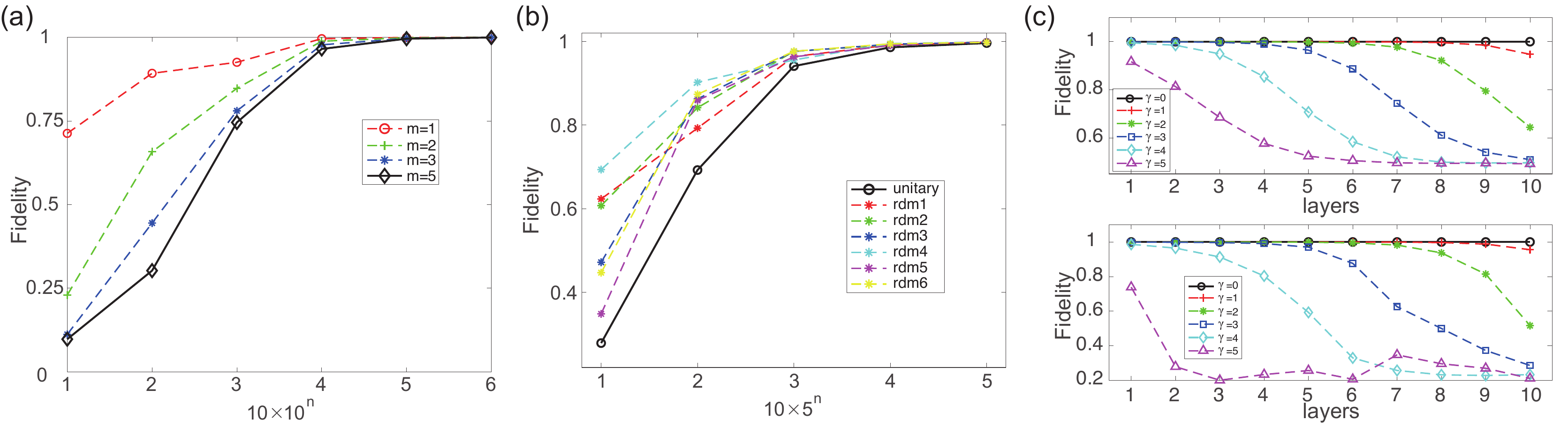}
  \caption{Numerical Simulation Results. (a) displays how the sampling size affects the fidelity of overlapping tomography for learning a single-layered 5-qubit system. Different sizes $m$ ($m \in \{1, 2, 3\}$) of the reduced density matrix are compared with the direct tomography of the whole system.
  (b) shows how the fidelity of different reduced density matrices $\text{rdm}_{i}$ ($i\in[1,6]$) varies with respect to the sampling size. Here, the size $m$ of Choi matrices is fixed as $m=2$.
  (c) presents the influence of quantum noise and layer depths on the accuracy of reconstructing multi-layered circuits. The upper (lower) sub-figure denotes a 2-qubit (2-qubit) principal system. The parameter $\gamma$ indicates the strength of the quantum noise.}
  \label{fig:simu}
\end{figure*}

First, we perform simulations to demonstrate that Algorithm~\ref{alg:paulitomo} can effectively gather complete information about the density matrix via overlapping tomography. We establish a 5-qubit system initialized as $\ket{0}^{\otimes 5}$ and perform the 5-qubit unitary operation according to Haar measure. We explore how the sampling size, denoted by $N = 10 \cdot 10^n$ ($n \in [1, 6]$), affects the accuracy of tomography, as illustrated in Fig.~\ref{fig:simu} (a). The accuracy of tomography is assessed by the fidelity between the reconstructed reduced-density matrix and its ideal counterpart. We also explore how the size of the reduced density matrix, denoted by $m$ ($m \in \{1, 2, 3\}$), influences tomographic accuracy. 
For instance, to obtain the average fidelity for the case $m=2$, we need to investigate $\binom{5}{2}=10$ different reduced density matrices.
As a comparison, the solid line labeled $m=5$ denotes direct tomography of the 5-qubit density matrix. A larger sampling size improves the accuracy of tomography.

Second, we conduct simulations to demonstrate the efficiency of Algorithm~\ref{alg:singlelayer} in reconstructing single-layer circuits. We set up a $4$-qubit principal system entangled with a $4$-qubit auxiliary system by the initial state $\ket{\Phi}^{\otimes 4}$. Due to space constraints, the details of the specific single-layer circuit under test are not provided here. We explore how the sampling size, denoted by $N = 10 \cdot 5^n$ ($n \in [1, 5]$), affects the accuracy of reconstructions, as illustrated in Fig.~\ref{fig:simu} (b). In each sampling, the chosen measurement bases and corresponding outcomes are recorded. Notice that the size $m$ of reduced density matrices remains fixed as $m=2$. Thus we need to investigate $\binom{4}{2}=6$ different reduced density matrices, indicated by the label $\text{rdm}_i$ in Fig.~\ref{fig:simu} (b). Similarly, the solid line labeled ``unitary" denotes the scenario involving unitaries comprising the whole single-layer circuit. Once again, the learning accuracy improves with a larger sampling size. If the accuracy exceeds a certain threshold $d_\mathcal{C}$ defined in Eq.~\eqref{eq:epsilon}, we can successfully identify the circuit structure and its configurations, as demonstrated in previous experiments.

Finally, we examine the influence of quantum noise on Algorithm~\ref{alg:learnmulti}. The quantum noise level is characterized by $\gamma\in[1, 5]$, introduced by incorporating random noise scaled by $5^{\gamma}\cdot 10^{-4}$ into the reconstructed reduced density matrices. Our simulations investigate a circuit with a maximum of $10$ layers generated from a continuous gate set. Consequently, it is not feasible to use distance metrics to infer gate configurations, and estimation errors will accumulate with each layer. Fig.~\ref{fig:simu} (c) demonstrates the accuracy of reconstructions concerning different layer depths, considering two distinct sizes $n=2$ or $4$ of the principal system. The solid line labeled $\gamma=0$ denotes the noiseless scenario. Naturally, the fidelity of tomography decreases as the number of learned layers increases. This highlights future research's significance in quantum noise and continuous gate sets in quantum circuit simulations and reconstructions.

\section*{Acknowledgment}
K. L. acknowledges Scientific Foundation for Youth Scholars of Shenzhen University and Guangdong Provincial Quantum Science Strategic Initiative (GDZX2303001, GDZX2403001). N. Y. acknowledges the support from University of Technology Sydney, Australia, where this research was conducted.

% \bibliographystyle{IEEEtran}
% \bibliography{main}


\end{document}